\documentclass[aps,pre,twocolumn,showpacs,floatfix]{revtex4-1}
\usepackage{graphicx}
\usepackage{dcolumn}
\usepackage{bm}
\usepackage{epsf}
\usepackage{subfigure}
\usepackage{epstopdf}
\usepackage{amsmath}
\usepackage{amssymb}
\usepackage[cspex,bbgreekl]{mathbbol}
\usepackage{color}

\begin{document}

\title{Stochastic thermodynamics under coarse-graining}

\author{Massimiliano Esposito}
\affiliation{Center for Nonlinear Phenomena and Complex Systems,
Universit\'e Libre de Bruxelles, CP 231, Campus Plaine, B-1050 Brussels, Belgium.}

\date{\today}

\begin{abstract}
A general formulation of stochastic thermodynamics is presented for open systems exchanging energy and particles with multiple reservoirs. By introducing a partition in terms of ``macrostates'' (e.g. sets of ``microstates"), the consequence on the thermodynamic description of the system is studied in detail. When microstates within macrostates rapidly thermalize, the entire structure of the microscopic theory is recovered at the macrostate level. This is not the case when these microstates remain out of equilibrium leading to additional contributions to the entropy balance. Some of our results are illustrated for a model of two coupled quantum dots.   
\end{abstract}

\maketitle

\section{Introduction}\label{Intro}

The last decade has witnessed major progress in nonequilibrium statistical mechanics. It is becoming increasingly clear that a consistent theory of nonequilibrium thermodynamics can be constructed for physical systems described by a stochastic Markovian dynamics. This so-called theory of stochastic thermodynamics generalizes the phenomenological formulations of nonequilibrium thermodynamics developed for systems locally close to equilibrium more than half a century ago \cite{PrigoThermo, PrigogineGlansdorff71, GrootMazur}. Early developments in this field where restricted to the ensemble averaged level and focused on steady state situations \cite{LuoVdBNicolis84, VanDenBroeckST86, Schnakenberg, NicoPrig77}. The crucial conceptual breakthrough came later and consisted in identifying the central thermodynamic quantities at the level of single stochastic trajectories \cite{Jarzynski97b, Kurchan98, Sekimoto98, Lebowitz99, Crooks98, Crooks99, Crooks00, Gaspard04, Seifert05, Peliti06, SeifertST08, AndrieuxGaspard07a, Sekimoto10, EspositoHarbola07PRE, EspoVdB10_Da, EspoVdB10_Db, Ge09, GeQian10}. The discovery of fluctuation theorems has played a major role in this regard \cite{Maes03, EspositoReview, HanggiFTRMP11, JarzynskiRev11, Harris07}. The second law of thermodynamics, traditionally expressed as an inequality at the ensemble averaged level, is now understood as resulting from a universal equality at the level of the full probability distribution of the entropy production defined at the trajectory level \cite{Seifert05, EspositoVdBPRL10}. These new theoretical developments are particularly important for the study of small systems subjected to sufficiently large and measurable fluctuations \cite{BustamanteRitortPD, QianPR12a, QianPR12b}. They coincided with an unprecedented development in the experimental techniques used to manipulate small systems and triggered a great deal of experimental studies in a variety of contexts such as single molecule stretching experiments \cite{BustamanteRitortPD, Bustamante02, Bustamante05}, nanomechanical oscillators work measurements \cite{Ciliberto05EPL}, spectroscopic measurement of trajectory entropies \cite{Seifert05exp, Seifert06exp, WoodsideNP11}, and electronic current fluctuations in full counting statistics experiments \cite{UtsumiSaitoFujisawaPRB10}. Since stochastic thermodynamics combines kinetics and thermodynamics it has also proved extremely useful to describe the finite-time thermodynamics (e.g. efficiency at finite power) of various nanodevices operating as thermodynamic machines \cite{GaveauSchulmanPRL10, SeifertPRL11, SeifertSchmiedlEPL08, EspositoPRL09, EspoKawLindVdB_PRE_10, EspoRuttCleuPRB, EspoLindVdB_EPL09_Dot, ChvostaMaass10, GaspardATP09}. Overall, this theory is becoming a fundamental tool for the study nanosciences.

The stochastic description underlying stochastic thermodynamics relies on a time-scale separation between system and reservoirs. The slow degrees of freedom entering the stochastic description constitute the ``system". They may be controlled by an external time-dependent force, but are also stochastically driven by hidden degrees of freedom which constitute the ``reservoirs". These later are so fast that they can be assumed to always remain at equilibrium. They can thus be characterized statistically by a temperature and a chemical potential. The systematic procedures to perform the elimination of these fast degrees of freedom (starting from a Hamiltonian description of the complete set of degrees of freedom) are nowadays well know \cite{Resibois77, KuboB98b, zwanzig, Breuer02}. While very fast equilibrated degrees of freedom constituting the reservoirs are ubiquitous at small scales (without them, the very existence of a thermodynamic description is compromised), nontrivial differences may exists between system degrees of freedom. This is particularly true in modeling biological systems where each level of description hides a significant underlying complexity. Although some of these systems degrees of freedom can be faster than others they may be maintained out-of-equilibrium thus preventing to treat them as reservoirs. Alternatively, one might only observe a subset of the true system degrees of freedom, since correctly identifying the system states is not always an easy task \cite{CluzelPRL06}. It is therefore important to understand how to formulate stochastic thermodynamics at a coarse grained level of description. This is the central topic of this paper. For convenience, we are going to call the true system states ``microstates" and states which lump together multiple true system states ``macrostates". While recent studies have investigated various aspects of such coarse graining at the level of the stochastic description \cite{Haseltine05, Peles06JCP, Vulpiani08, Andrieux11, QianSantillanPRE11}, as well as some of its implications at the level of the thermodynamics description \cite{VandenBroeck08, SeifertAmannJCP10, Nicolis11, Seifert11EPJE}, the present paper analyses for the first time the effect of coarse-graining on stochastic thermodynamics as a whole. A rewarding outcome of this study is that the coarse graining procedure a posteriori unambiguously clarifies the implicit assumptions made when formulating stochastic thermodynamics at the microscopic level. 

This paper is organized as follows. A general formulation of stochastic thermodynamics for open systems is presented in section \ref{Basics}. The case of multiple reservoirs as well as the limit of a single reservoir is considered. In the later case standard equilibrium thermodynamics is recovered in the reversible limit. In section \ref{CG}, the coarse-graining procedure is applied to the dynamics and a natural approximation scheme is presented for situations where the partition is motivated by a time scale separation between the micro and macro-level. In section \ref{EntBalSec}, the effect of coarse-graining on the entropy balance is discussed for various scenarios. In section \ref{Applic}, applications to a double quantum dot model are presented. Conclusions are drawn in section \ref{Conc}.

\section{Stochastic thermodynamics in the grand canonical ensemble}\label{Basics}

\subsection{Multiple reservoirs}

We consider a system described by a set of states $i$ with a given system energy $\epsilon_i$, number of particles $n_i$, and equilibrium entropy $s_i$. Transitions between these states are induced by multiple reservoirs $\nu$ with a given chemical potential $\mu^{(\nu)}$ and temperature $T^{(\nu)}$. Each state has therefore a given grand potential (or Landau potential) $\omega_i^{(\nu)}$ with regards to the reservoir $\nu$
\begin{eqnarray}
\omega_i^{(\nu)} = \epsilon_i - \mu^{(\nu)} n_i - T^{(\nu)} s_i . \label{GrandPotential}
\end{eqnarray}
We assume that the system energy, number of particle, and entropy of a level $i$, as well as the reservoirs chemical potentials and temperatures (i.e. all terms in $\omega_i^{(\nu)}$) may be controlled in a time dependent manner by an external agent. We will refer to this process as external driving. Without loss of generality, we parametrized this time dependence through $\lambda$ so that $\dot{\omega}_i^{(\nu)}=\dot{\lambda} \; \partial_{\lambda} \omega_i^{(\nu)}$. The dynamics resulting from the stochastic transitions between system states is ruled by the Makovian master equation  
\begin{eqnarray}
\dot{p}_i = \sum_{j} W_{ij} p_{j} . \label{MasterEq}
\end{eqnarray}
The rate matrix, which may depend on time due to the external driving, satisfies $\sum_{i} W_{ij}=0$ and is assumed irreducible. The unique stationary distribution $p_{i}^{\rm st}$ is thus obtained by solving $\sum_{j} W_{ij} p_{j}^{\rm st}=0$. Since transitions can be due to different reservoirs, the rate matrix is decomposed in their respective contribution 
\begin{eqnarray}
W_{ij} = \sum_{\nu} W_{ij}^{(\nu)} .\label{AddRateMat}
\end{eqnarray}
Because reservoirs are assumed to always remain at equilibrium, the rate matrix satisfies {\it local detailed balance}  
\begin{eqnarray}
\frac{W_{ij}^{(\nu)}}{W_{ji}^{(\nu)}} = \exp{\bigg(-\frac{\omega_i^{(\nu)} - \omega_j^{(\nu)}}{k_b T^{(\nu)}}\bigg)}, \label{LocalDB}
\end{eqnarray}
where $k_b$ is the Boltzmann constant. This property guaranties that a system in contact with a single reservoir (or equivalently in contact with multiple reservoirs with identical temperatures and chemical potentials) and in absence of external driving will eventually reach the grand canonical equilibrium  
\begin{eqnarray}
p_i^{\rm eq} = \exp{\bigg(-\frac{\omega_i-\Omega^{\rm eq}}{k_b T}\bigg)} . \label{GrandCanDist}
\end{eqnarray}
This distribution satisfies the {\it detailed balance} condition 
\begin{eqnarray}
W_{i j}^{(\nu)} p_{j}^{\rm eq} = W_{j i}^{(\nu)} p_{i}^{\rm eq} \ \ , \ \ \forall \; \nu,i,j  \label{DBtotal}
\end{eqnarray}
which defines equilibrium and indicates that all currents vanish in the system.

The system energy and number of particle are naturally given by the ensemble average
\begin{eqnarray}
E = \sum_i \epsilon_i p_i \ \ \;, \ \ N = \sum_i n_i p_i. \label{EnergyMatter}
\end{eqnarray}
However, the system entropy is not simply the ensemble average of the entropy of each state $i$ but also contains an information (Shanon-like) contribution  
\begin{eqnarray}
S = \sum_i [ s_i - k_b \ln p_i ] p_i \label{Entropy} .
\end{eqnarray}
We note that this defines entropy out-of-equilibrium.

The change in energy and number of particle can be expressed as
\begin{eqnarray}
\dot{E} &=& \sum_i \epsilon_i \dot{p}_i+\sum_i \dot{\epsilon}_i p_i = \sum_{\nu} I^{(\nu)}_E + \dot{\lambda} \; \partial_{\lambda} E \label{EnergyChange}\\
\dot{N} &=& \sum_i n_i \dot{p}_i+\sum_i \dot{n}_i p_i = \sum_{\nu} I^{(\nu)}_N + \dot{\lambda} \; \partial_{\lambda} N \label{MatterChange}.
\end{eqnarray}
The second contribution is due to the external driving while the first is due to the reservoirs and is expressed in terms of the energy and matter currents entering the system  
\begin{eqnarray}
&&I^{(\nu)}_E = \sum_{i,j} W_{ij}^{(\nu)} p_{j} (\epsilon_i-\epsilon_j) \label{EnergyCurrent}\\
&&I^{(\nu)}_N = \sum_{i,j} W_{ij}^{(\nu)} p_{j} (n_i-n_j) .\label{MatterCurrent}
\end{eqnarray}
The change in the system entropy, 
\begin{eqnarray}
\dot{S} = \sum_i [ s_i - k_b \ln p_i ] \dot{p}_i +  \sum_i \dot{s}_i p_i \label{EntropyBalance} ,
\end{eqnarray}
can be decomposed in analogy with irreversible thermodynamics as \cite{PrigoThermo,GrootMazur}
\begin{eqnarray}
\dot{S} = \dot{S}_{\rm \bold i} + \dot{S}_{\rm \bold e}. \label{EntropyBalanceBis}
\end{eqnarray}
The non-negative entropy production is given by
\begin{eqnarray}
\dot{S}_{\rm \bold i} = k_b \sum_{\nu,i,j} W_{ij}^{(\nu)} p_{j} \ln \frac{W_{ij}^{(\nu)} p_{j}}{W_{ji}^{(\nu)} p_{i}} \geq 0 .\label{EntropyProd}
\end{eqnarray}
Non-negativity is proved using $\ln x \leq x-1$. Entropy production only vanish for reversible transformations, i.e. transformation along which the detailed balance condition (\ref{DBtotal}) is satisfied. The entropy flow in turn is given by
\begin{eqnarray}
\dot{S}_{\rm \bold e} 
&=& \sum_{\nu,i,j} W_{ij}^{(\nu)} p_{j} \big( s_i -k_b \ln \frac{W_{ij}^{(\nu)}}{W_{ji}^{(\nu)}} \big) + \sum_i \dot{s}_i p_i \label{EntropyFlow}\\
&=& \sum_{\nu} \frac{\dot{{\cal Q}}^{(\nu)}}{T^{(\nu)}} + \dot{\lambda} \; \partial_{\lambda} S. \nonumber
\end{eqnarray}
The second contribution to the entropy flow is due to the external driving which reversibly modifies the equilibrium entropy associated to the internal structure of the states. The first contribution is due to the reservoirs and is expressed in term of the heat flowing from reservoir $\nu$ to the system
\begin{eqnarray}
\dot{{\cal Q}}^{(\nu)} = I^{(\nu)}_E-\mu^{(\nu)} I^{(\nu)}_M . \label{Heat}
\end{eqnarray}
As a result, work reads 
\begin{eqnarray}
\dot{{\cal W}} = \dot{\lambda} \; \partial_{\lambda} E + \sum_{\nu} \mu^{(\nu)} I^{(\nu)}_M \label{Work}
\end{eqnarray}
and the first law is satisfied
\begin{eqnarray}
\dot{E} = \dot{{\cal W}} + \sum_{\nu} \dot{{\cal Q}}^{(\nu)}. \label{FirstPrinciple}
\end{eqnarray}
We note that some authors do not incorporate the particle current contribution in the definition of heat and work and keep it as a distinct contribution to the entropy flow.

\subsection{Single reservoir} \label{1resStochThermo}

We now show the simplifications which take place when the system interacts with a single reservoir (from now on, we stop repeating that single reservoir also refers to multiple reservoirs with identical thermodynamic properties). Using (\ref{EntropyFlow}) with (\ref{LocalDB}), the entropy flow becomes
\begin{eqnarray}
T \dot{S}_{\rm \bold e} = \sum_{i} (\epsilon_i-\mu n_i) \dot{p}_i + \sum_i \dot{s}_i p_i = \dot{{\cal Q}}+ \dot{\lambda} \; \partial_{\lambda} S , \label{EntFlow1res}
\end{eqnarray}
where heat can be written, using (\ref{Heat}), (\ref{EnergyChange}) and (\ref{MatterChange}), as
\begin{eqnarray}
\dot{{\cal Q}} = (\dot{E}-\mu \dot{N}) - \dot{\lambda} (\partial_{\lambda} E - \mu \partial_{\lambda} N) \label{Heat1res}.
\end{eqnarray}
Using (\ref{EntropyBalance}) with (\ref{EntFlow1res}), entropy production now reads
\begin{eqnarray}
T \dot{S}_{\rm \bold i} = - \sum_i \big( \omega_i + k_b T \ln p_i \big) \dot{p}_i \geq 0  \label{EntProd1res}.
\end{eqnarray}
Introducing the system nonequilibrium grand potential (or Landau potential)
\begin{eqnarray}
\Omega = E - \mu N - T S = \sum_i \big( \omega_i + k_b T \ln p_i \big) p_i , \label{SysGrandPotential}
\end{eqnarray}
we find that entropy production can be expressed as
\begin{eqnarray}
T \dot{S}_{\rm \bold i} = -(\dot{\Omega} -\dot{\lambda} \; \partial_{\lambda} {\Omega}) \geq 0  \label{EntProd1resBis}.
\end{eqnarray}
If we assume that $\partial_{\lambda} n_i=\partial_{\lambda} s_i=0$ as is often the case, we find that (\ref{EntProd1resBis}) reduces to
\begin{eqnarray}
T \dot{S}_{\rm \bold i} = - \big( \dot{E} - \dot{\lambda} \; \partial_{\lambda} {E} - \mu \dot{N} - T \dot{S} \big) \geq 0  \label{EntProd1resTris}.
\end{eqnarray}

In absence of external driving, a system prepared in an arbitrary initial nonequilibrium state will always relax to equilibrium where all quantities stop evolving (i.e. $\dot{S}_{\rm \bold i}=\dot{S}_{\rm \bold e}=\dot{S}=0$ and $\dot{\Omega}=\dot{E}=\dot{N}=0$). Their stationary value is given by $X|_{\rm eq}$, where $X=\Omega,E,N,S$ and $|_{\rm eq}$ denotes that $p_i$ is replaced by $p_i^{\rm eq}$ given by (\ref{GrandCanDist}). Note that using (\ref{SysGrandPotential}), we verify that $\Omega |_{\rm eq} = \Omega^{\rm eq}$.
We now consider a system initially at equilibrium and subjected to a slow (compared to the typical relaxation time of the system) external driving. Its probability distribution $p_i$ will follow the instantaneous equilibrium grand canonical distribution $p_i^{\rm eq}$ and the grand potential, the average energy and number of particles, and entropy, all become state functions
\begin{eqnarray}
\dot{\lambda} (\partial_{\lambda} X)|_{\rm eq} = \dot{X}|_{\rm eq} \  \ {\rm for} \  \ X=\Omega,E,N,S.
\end{eqnarray}
Such quasi-static transformations are called reversible because entropy production remains zero all along the process $\dot{S}_{\rm \bold i}|_{\rm eq}=0$. Consequently, the changes in the system entropy are given by the entropy flow
\begin{eqnarray}
\dot{S}|_{\rm eq}=\dot{S}_{\rm \bold e}|_{\rm eq} =\frac{\dot{{\cal Q}}}{T}|_{\rm eq} + \dot{\lambda} \; (\partial_{\lambda} S)|_{\rm eq}.
\end{eqnarray}
We note that in this case, we recover the fundamental equation of equilibrium thermodynamics from (\ref{EntProd1resTris}). This shows how traditional equilibrium thermodynamics is recovered from stochastic thermodynamics in the reversible limit. 

\section{Coarse graining} \label{CG}

We now define a set of ``macrostates" denoted by $k$ and assume that each ``microstates" $i$ leads to a unique macrostate $k=k(i)$. This terminology is used for convenience and does not necessarily refer to a notion of size. We use the compact notation $i_k$ to denote microstates which lead to the macrostate $k$. The probability to find the system in a macrostate $k$ is given by
\begin{eqnarray}
P_k = \sum_{i_k} p_{i_k} = \sum_{i} \delta_{\rm Kr}[k-k(i)] p_i \label{MacroProb}
\end{eqnarray}
The conditional probability to be in the microstate $i_k$ being in the macrostate $k$ is denoted
\begin{eqnarray}
\mathbb{P}_{i_k} = p_{i_k}/P_k \label{CondProb}
\end{eqnarray}
We verify that $\sum_{i_k} \mathbb{P}_{i_k} = 1$.

\subsection{Dynamics}

Writing the master equation (\ref{MasterEq}) in terms of (\ref{CondProb}), we find 
\begin{eqnarray}
\dot{P}_k \mathbb{P}_{i_k} + P_{k} \dot{\mathbb{P}}_{i_k} = \sum_{k'} P_{k'} \sum_{\nu,j_{k'}} W_{i_k j_{k'}}^{(\nu)} \mathbb{P}_{j_{k'}} \label{ComplDyn}.
\end{eqnarray}
Summing this equation over $i_k$, we find a master equation ruling the dynamics of the macrostates
\begin{eqnarray}
\dot{P}_k = \sum_{\nu,k'} V_{kk'}^{(\nu)} P_{k'} \label{MEmacrostates}.
\end{eqnarray}
This equation is not closed because the macroscopic rate matrix,
\begin{eqnarray}
V_{kk'}^{(\nu)}= \sum_{i_k,j_{k'}} W_{i_k j_{k'}}^{(\nu)} \mathbb{P}_{j_{k'}} \label{CGRate},
\end{eqnarray}
depends on the dynamics of the microstates through $\mathbb{P}_{j_{k'}}$. We verify that $\sum_k V_{kk'} = 0$. In general, even for a time-independent microscopic rate matrix, as long as the distribution of the microlevels evolves (i.e. $\mathbb{P}_{i_k}$ is time dependent), $V_{kk'}$ will be time dependent. 

\subsection{Time scale separation} \label{AdiabElim}

We now assume that the coarse graining procedure is chosen such that the dynamics between microstates belonging to the same macrostate is much faster then between microstates belonging to different macrostates. In other words, the rate matrix is such that $W_{i_k j_k} \gg W_{i_k j_{k'}}$ for $k \neq k'$. Therefore the results presented in this subsection can be proved using perturbation theory as shown in appendix \ref{AppAdiabElimRigor}. In such situations, the conditional probabilities $\mathbb{P}_{i_k}$ evolve much faster then the macrostates probabilities $P_k$. On short time scales, denoted $\tau_{\rm mic}$, the macrostates probabilities $P_k$ barely change while the $\mathbb{P}_{i_k}$'s obey an almost isolated dynamics inside the macrostates $k$ eventually relaxing to the stationary distribution $\mathbb{P}^{\rm st}_{j_k}$ defined by  
\begin{eqnarray}
\sum_{j_k} W_{i_k j_k} \mathbb{P}^{\rm st}_{j_k} = 0 \label{ApproxDyn}.
\end{eqnarray}

If the transitions between microstates belonging to a given macrostate $k$ are due to a single reservoir, due to local detailed balance property (\ref{LocalDB}), $\mathbb{P}^{\rm st}_{i_k}$ will be given by the equilibrium distribution
\begin{eqnarray}
\mathbb{P}_{i_k}^{\rm eq}=\exp{\bigg(- \frac{\omega_{i_k}-\mathbb{\Omega}^{\rm eq}(k)}{k_b T}\bigg)} \label{LocalEqui},
\end{eqnarray}
and all currents within the macrostate vanish (i.e. detailed balance is satisfied within $k$)
\begin{eqnarray}
W_{i_k j_k} \mathbb{P}_{j_k}^{\rm eq} = W_{j_k i_k} \mathbb{P}_{i_k}^{\rm eq} .\label{DBink}
\end{eqnarray}

Turning back to the general case of multiple reservoirs, for times much longer than $\tau_{\rm mic}$, the $P_k$'s will start evolving following the approximate macrostate dynamics (\ref{MEmacrostates})
\begin{eqnarray}
\dot{P}_k = \sum_{\nu,k'} V_{kk'}^{(\nu) \rm st} P_{k'} \label{MEmacrostatesApprox} .
\end{eqnarray}
This equation is closed because, thanks to the time scale separation, the exact macroscopic rate matrix can be approximated by 
\begin{eqnarray}
V_{kk'}^{(\nu) \rm st}= \sum_{i_k,j_{k'}} W_{i_k j_{k'}}^{(\nu)} \mathbb{P}^{\rm st}_{j_{k'}} . \label{CGRateApprox}
\end{eqnarray}
Over a characteristic time $\tau_{\rm mac}$, the $P_k$'s will also reach a stationary distribution $P_{k}^{\rm st}$ defined by
\begin{eqnarray}
\sum_{\nu,k'} V_{kk'}^{(\nu) \rm st} P_{k'}^{\rm st} =0 \label{MEmacrostatesApproxSt} .
\end{eqnarray}

If the entire system is in contact with a single reservoir, the macroscopic rate matrix $V_{kk'}^{(\nu) \rm st}$ becomes $V_{kk'}^{\rm eq}$, meaning that $\mathbb{P}^{\rm st}_{i_k}$ is replaced by $\mathbb{P}^{\rm eq}_{i_k}$ in (\ref{CGRateApprox}). In this case, using (\ref{LocalDB}) and (\ref{CGRate}), we recover the property of local detailed balance at the level of the macroscopic rates 
\begin{eqnarray}
\frac{V_{kk'}^{\rm eq}}{V_{k'k}^{\rm eq}} = \exp{\bigg(- \frac{\mathbb{\Omega}^{\rm eq}(k)-\mathbb{\Omega}^{\rm eq}(k')}{k_b T} \bigg)}. \label{MacroLDB}
\end{eqnarray}
As a result, the stationary distribution of the macroscopic states $P_{k}^{\rm st}$ is given by the equilibrium distribution 
\begin{eqnarray}
P_k^{\rm eq} = \exp{\bigg(- \frac{\mathbb{\Omega}^{\rm eq}(k)-\Omega^{\rm eq}}{k_b T}\bigg)} . \label{MacProbEq}
\end{eqnarray}
Using (\ref{GrandCanDist}), (\ref{LocalEqui}) and (\ref{MacProbEq}), we verify that 
\begin{eqnarray}
p_{i_k}^{\rm eq} = P_k^{\rm eq} \mathbb{P}_{i_k}^{\rm eq}.
\end{eqnarray}
This means that over times larger than $\tau_{\rm mac}$ the full system reaches equilibrium and detailed balance condition (\ref{DBtotal}) is satisfied. 

\subsection{Energy and entropy}\label{EnergyMatterSec}

We now turn to the effect of a general coarse-graining on the energy, number of particles and entropy. The system energy and particle number (\ref{EnergyMatter}) can be expressed as 
\begin{eqnarray}
E=\sum_{k} \mathbb{E}(k) P_k \ \ , \ \ N= \sum_{k} \mathbb{N}(k) P_k  ,\label{ObsMeso1}
\end{eqnarray}
where the average energy and number of particle conditional on being on a macrostate $k$ are given by
\begin{eqnarray}
\mathbb{E}(k) = \sum_{i_k} \epsilon_{i_k} \mathbb{P}_{i_k} \ \ , \ \ \mathbb{N}(k) = \sum_{i_k} n_{i_k} \mathbb{P}_{i_k} \label{ObsMeso1b} .
\end{eqnarray}
Their evolution can be expressed as 
\begin{eqnarray}
\dot{E} &=& \sum_{k} \mathbb{E}(k) \dot{P}_k + \sum_{k} \dot{\mathbb{E}}(k) P_k \label{ObsDotMesoEnergy} \\
\dot{N} &=& \sum_{k} \mathbb{N}(k) \dot{P}_k + \sum_{k} \dot{\mathbb{N}}(k) P_k \label{ObsDotMesoMatter}.
\end{eqnarray}
The system entropy (\ref{Entropy}) can be rewritten as
\begin{eqnarray}
S = \sum_k \big( \mathbb{S}(k) - k_b \ln P_k \big) P_k , \label{EntMeso1} 
\end{eqnarray}
where the entropy conditional on being on a macrostate $k$ is given by
\begin{eqnarray}
\mathbb{S}(k) = \sum_{i_k} \big( s_{i_k} - k_b \ln \mathbb{P}_{i_k} \big) \mathbb{P}_{i_k} \label{EntMeso2}.
\end{eqnarray}
The entropy evolution reads 
\begin{eqnarray}
\dot{S} = \sum_k \big( \mathbb{S}(k) - k_b \ln P_k \big) \dot{P_k} + \sum_k \dot{\mathbb{S}}(k) P_k \label{EntMeso1Evol} ,
\end{eqnarray}
where 
\begin{eqnarray}
\dot{\mathbb{S}}(k) = \sum_{i_k} \big( s_{i_k} - k_b \ln \mathbb{P}_{i_k} \big) \dot{\mathbb{P}}_{i_k} + \sum_{i_k} \dot{s}_{i_k} \mathbb{P}_{i_k} \label{EntMeso2Evol}.
\end{eqnarray}
We note that the evolution of energy, number of particles, and entropy, expressed in terms of the macrostates [(\ref{ObsDotMesoEnergy}), (\ref{ObsDotMesoMatter}) and (\ref{EntMeso1Evol})] has the same form as the original evolution expressed in terms of microstates [(\ref{EnergyChange}), (\ref{MatterChange}) and (\ref{EntropyBalance})]. The key difference is that the evolution of the quantities defined on the macrostates is not anymore exclusively due to the external driving, but also contains the internal dynamics of the macrostate. Remarkably however, the form of this internal dynamics expressed in term of conditional probabilities, $\mathbb{P}_{i_k}$, is also the same as the original evolution expressed in terms of microstates. 

\section{Entropy balance}\label{EntBalSec}

\subsection{Single reservoir}

We now formulate the entropy balance for the case of a single reservoir. 
Using (\ref{CondProb}) and (\ref{ComplDyn}) in (\ref{EntFlow1res}) and (\ref{Heat1res}), the entropy flow can be rewritten as
\begin{eqnarray}
T \dot{S}_{\rm \bold e} = \dot{{\cal Q}} + \dot{\lambda} \; T \sum_k \partial_{\lambda} \mathbb{S}(k) P_k \label{EntFlowMeso1res},
\end{eqnarray}
where heat is given by
\begin{eqnarray}
\dot{{\cal Q}} &=& \sum_k \big( \mathbb{E}(k) - \mu \mathbb{N}(k) \big) \dot{P}_k \label{HeatMeso1res} \\
&& + \sum_k \bigg( \dot{\mathbb{E}}(k) - \mu \dot{\mathbb{N}}(k) -\dot{\lambda} \big( \partial_{\lambda} \mathbb{E}(k) - \mu \partial_{\lambda} \mathbb{N}(k) \big) \bigg) P_k \nonumber .
\end{eqnarray}
It is worth noting that 
\begin{eqnarray} 
&& \dot{\mathbb{E}}(k) - \mu \dot{\mathbb{N}}(k) -\dot{\lambda} \big( \partial_{\lambda} \mathbb{E}(k) - \mu \partial_{\lambda} \mathbb{N}(k)\big) \\
&&\hspace{4cm} = \sum_{i_k} \big( \epsilon_{i_k} + \mu n_{i_k} \big) \dot{\mathbb{P}}_{i_k} . \nonumber
\end{eqnarray}
The form taken by the entropy flow at the macrostate level contains two types of contributions. The first is made of the second term of (\ref{EntFlowMeso1res}) and the first term of (\ref{HeatMeso1res}). It has the exact same form as the entropy flow at the microstates level (\ref{EntFlow1res}) with (\ref{Heat1res}). The second contribution consists in an ensemble average over the macrostates probabilities, $P_k$, of the heat flow within each macrostate. We now turn to entropy production. Using the system entropy (\ref{EntMeso1Evol}), we find that entropy production (\ref{EntProd1res}) can be written as 
\begin{eqnarray}
T \dot{S}_{\rm \bold i} &=& - \sum_{k} \big( \mathbb{\Omega}(k) + k_b T \ln P_k \big) \dot{P}_k \label{EntProdMeso2} \\
&&- \sum_{k} \big( \dot{\mathbb{\Omega}}(k) - \dot{\lambda} \; \partial_{\lambda} \mathbb{\Omega}(k) \big) P_k \nonumber .
\end{eqnarray}
We introduced the macrostate grand potential
\begin{eqnarray}
\mathbb{\Omega}(k) &=& \mathbb{E}(k) - \mu \mathbb{N}(k) - T \mathbb{S}(k) \label{LocalNonEquiGrandPot}\\
&=& \sum_{i_k} \big( \omega_{i_k} + k_b T \ln \mathbb{P}_{i_k} \big) \mathbb{P}_{i_k} \nonumber ,
\end{eqnarray}
which is connected to the grand potential (\ref{SysGrandPotential}) by
\begin{eqnarray}
\Omega = \sum_k \big( \mathbb{\Omega}(k) + k_b T \ln P_k \big) P_k \label{DefXk} .
\end{eqnarray}
It is worth noting that 
\begin{eqnarray} 
\dot{\mathbb{\Omega}}(k) - \dot{\lambda} \; \partial_{\lambda} \mathbb{\Omega}(k) 
= \sum_{i_k} \big( \Omega_{i_k} + k_b T \ln \mathbb{P}_{i_k} \big) \dot{\mathbb{P}}_{i_k}.
\end{eqnarray}
As for the entropy flow, the form taken by the entropy production at the macrostate level contains two types of contributions. The first term in (\ref{EntProdMeso2}) has the exact same form as the entropy production at the microstates level (\ref{EntProd1res}), and the second term is an ensemble average over the macrostates probabilities, $P_k$, of the entropy production arising from within each macrostate.

We now turn to the situation described in \ref{AdiabElim} where microvariables evolve faster than macrovariables. We consider the system evolution over timescales longer than $\tau_{\rm mic}$. We assume that the external driving is sufficiently slow to keep the microstates within macrostates at equilibrium, i.e. $\mathbb{P}_{i_k}$ is replaced by $\mathbb{P}^{\rm eq}_{i_k}$ given by (\ref{LocalEqui}). It can however be fast enough to keep the macrostate probabilities, $P_k$, far from equilibrium. As a result, using (\ref{LocalNonEquiGrandPot}), we verify that
\begin{eqnarray}
\mathbb{\Omega}(k)\vert_{\rm eq} = \mathbb{\Omega}^{\rm eq}(k) \label{LocalEquiGrandPot}
\end{eqnarray}
where $\vert_{\rm eq}$ means that $\mathbb{P}_{i_k}$ in the expression has to be replaced by $\mathbb{P}^{\rm eq}_{i_k}$. Defining 
\begin{eqnarray}
\mathbb{X}^{\rm eq}(k) \equiv \mathbb{X}(k) \vert_{\rm eq} \  \ {\rm for} \  \ \mathbb{X}=\mathbb{\Omega},\mathbb{E},\mathbb{N},\mathbb{S},
\end{eqnarray}
we find the important property 
\begin{eqnarray}
\dot{\lambda} \; \partial_{\lambda} \mathbb{X}^{\rm eq}(k) = \dot{\mathbb{X}}(k) \vert_{\rm eq}  \  \ {\rm for} \  \ \mathbb{X}=\mathbb{\Omega},\mathbb{E},\mathbb{N},\mathbb{S}  \label{EvolGrandPotonk} ,
\end{eqnarray}
which translates the fact that the microstates within macrostates evolve reversibly.
As a result, using (\ref{EntMeso1Evol}), the system entropy evolves as
\begin{eqnarray}
\hspace{-0.2cm}
\dot{S} = \sum_k \big( \mathbb{S}^{\rm eq}(k) - k_b \ln P_k \big) \dot{P_k} + \dot{\lambda} \sum_k \partial_{\lambda} \mathbb{S}^{\rm eq}(k) P_k \label{EntMeso1EvolEq} .
\end{eqnarray}
The entropy flow, using (\ref{EvolGrandPotonk}) with (\ref{EntFlowMeso1res}) and (\ref{HeatMeso1res}), becomes
\begin{eqnarray}
T \dot{S}_{\rm \bold e} = \dot{{\cal Q}} + \dot{\lambda} \sum_k \partial_{\lambda} \mathbb{S}^{\rm eq}(k) P_k , \label{EntFlow1resMicroEq}
\end{eqnarray}
where heat is given by 
\begin{eqnarray}
\dot{{\cal Q}} = \sum_k \big( \mathbb{E}^{\rm eq}(k) - \mu \mathbb{N}^{\rm eq}(k) \big) \dot{P}_k .\label{Heat1resMicroEq}
\end{eqnarray}
Entropy production, using (\ref{EvolGrandPotonk}) with (\ref{EntProdMeso2}) and (\ref{EvolGrandPotonk}), reads
\begin{eqnarray}
T \dot{S}_{\rm \bold i} = - \sum_k \big( \mathbb{\Omega}^{\rm eq}(k) + k_b T \ln P_k \big) \dot{P}_k  \label{EntProd1resMicroEq}.
\end{eqnarray}
We notice that the second term in (\ref{HeatMeso1res}) as well as in (\ref{EntProdMeso2}), which both arise from the dynamics within the macrostates, have vanished due to (\ref{EvolGrandPotonk}). Using the local detailed balance property of the macroscopic rates (\ref{MacroLDB}), the entropy flow (\ref{EntFlow1resMicroEq}) can finally be rewritten as
\begin{eqnarray}
\dot{S}_{\rm \bold e} &=& \sum_{k,k'} V_{kk'}^{\rm eq} P_{k'} \big(\mathbb{S}^{\rm eq}(k) - k_b \ln \frac{V_{kk'}^{\rm eq}}{V_{k'k}^{\rm eq}} \big) \label{EntflowMeso1EvolEq} \\
&&+ \dot{\lambda} \sum_k \partial_{\lambda} \mathbb{S}^{\rm eq}(k) P_k \nonumber
\end{eqnarray}
and the entropy production (\ref{EntProd1resMicroEq}) as 
\begin{eqnarray}
\dot{S}_{\rm \bold i} = k_b \sum_{k,k'} V_{kk'}^{\rm eq} P_{k'} \ln \frac{V_{kk'}^{\rm eq}P_{k'}}{V_{k'k}^{\rm eq}P_{k}} \geq 0 .\label{EntprodMeso1EvolEq}
\end{eqnarray}
This clearly shows that when microstates within macrostates are at equilibrium, stochastic thermodynamics assumes the same form at the macrostate level as at the microstate level. This can be clearly seen when comparing (\ref{EntProd1resMicroEq}) with (\ref{EntProd1res}) or (\ref{EntprodMeso1EvolEq}) with (\ref{EntropyProd}) as well as when comparing (\ref{EntFlow1resMicroEq}) with (\ref{EntFlow1res}) or (\ref{EntflowMeso1EvolEq}) with (\ref{EntropyFlow})]. This important result demonstrates a posteriori that the theory of stochastic thermodynamics makes a key assumption at its most fundamental level of description described in section \ref{Basics}: the internal structure of the states entering the stochastic description may evolve due to external driving but always does so reversibly, i.e. by remaining at equilibrium. In next section, we will generalize this result to macrostates in contact with multiple reservoirs. 

All terms containing $\dot{\lambda}$ vanish in absence of driving. However, the terms containing $\dot{\lambda}$ at the level of the internal dynamics of macrostates [e.g. in (\ref{EntMeso1EvolEq}) and (\ref{EntFlow1resMicroEq})] may also be omitted for driven systems when the external driving can be assumed to act similarly on all microstates belonging to the same macrostate.

\subsection{Multiple reservoirs}\label{MesoMutiResSec}

We now consider the general case of different reservoirs. We start by separating the evolution of the system entropy (\ref{EntMeso1}) in three contributions
\begin{eqnarray}
\dot{S} = \dot{S}^{(1)} + \dot{S}^{(2)} + \dot{S}^{(3)} \label{EntDotMeso}.
\end{eqnarray}
The first is the evolution of the (Shannon) information entropy expressed in terms of the macrostates probabilities. It corresponds to the entropy evolution of a system made of macrostates without internal structure   
\begin{eqnarray}
\dot{S}^{(1)} = - k_b \sum_k \dot{P_k} \ln P_k = \frac{d}{dt} (- k_b \sum_k P_k \ln P_k) \label{EntManyResSplit1} .
\end{eqnarray}
The second contribution is an average over the ensemble of macrostates of the entropy change occurring within each macrostate
\begin{eqnarray}
\dot{S}^{(2)} &=& \sum_k P_{k} \bigg( \sum_{i_k} \dot{s}_{i_k} \mathbb{P}_{i_k} + \sum_{\nu,i_k,j_k} W_{i_k,j_k}^{(\nu)} \mathbb{P}_{j_k} \label{EntManyResSplit2}\\
&&\hspace{2cm}\times \big(s_{i_k}-s_{j_k} - k_b \ln \mathbb{P}_{i_k}/\mathbb{P}_{j_k} \big) \bigg) \nonumber .
\end{eqnarray}
The remaining third contribution contains the entropy changes due to transitions between microstates belonging to different macrostates 
\begin{eqnarray}
\dot{S}^{(3)} &=& \sum_{k,k'(\neq k)} P_{k'} \bigg( \sum_{\nu,i_k,j_{k'}} W_{i_k,j_{k'}}^{(\nu)} \mathbb{P}_{j_{k'}} \label{EntManyResSplit3}\\
&&\hspace{1.8cm}\times \big(s_{i_k}-s_{j_{k'}} - k_b \ln \mathbb{P}_{i_k}/\mathbb{P}_{j_{k'}} \big) \bigg) \nonumber.
\end{eqnarray}
We note that the second and third contribution sum up to
\begin{eqnarray}
\dot{S}^{(2)}+ \dot{S}^{(3)} = \frac{d}{dt} (\sum_{k} \mathbb{S}(k) P_k)  \label{EntManyResSplit2+3}.
\end{eqnarray}

We proceed with the entropy production (\ref{EntropyProd}) which we also split in three parts 
\begin{eqnarray}
\dot{S}_{\rm \bold i} = \dot{S}^{(1)}_{\rm \bold i} + \dot{S}^{(2)}_{\rm \bold i} + \dot{S}^{(3)}_{\rm \bold i} \geq 0 \label{EP}.
\end{eqnarray}
The first contribution is the entropy production that one would write for macrostates probabilities obeying a closed Markovian dynamics
\begin{eqnarray}
\dot{S}^{(1)}_{\rm \bold i} = k_b \sum_{\nu,k,k'} V_{kk'}^{(\nu)} P_{k'} \ln \frac{V_{kk'}^{(\nu)} P_{k'}}{V_{k'k}^{(\nu)} P_{k}} .\label{EP1}
\end{eqnarray}
The second one is an ensemble average over the macrostates of the entropy production arising from the dynamics within macrostates 
\begin{eqnarray}
\dot{S}^{(2)}_{\rm \bold i} = k_b \sum_{k} P_{k} \bigg( \sum_{\nu,i_k,j_k} W_{i_k,j_k}^{(\nu)} \mathbb{P}_{j_k}  
\ln \frac{W_{i_k,j_k}^{(\nu)} \mathbb{P}_{j_k}}{W_{j_k,i_k}^{(\nu)} \mathbb{P}_{i_k}} \bigg) .\label{EP2}
\end{eqnarray}
The remaining contribution can be written as
\begin{eqnarray}
\dot{S}^{(3)}_{\rm \bold i} = \sum_{\nu,k,k'(\neq k)} V_{kk'}^{(\nu)} P_{k'} \mathbb{D}_{kk'}^{(\nu)} \label{EP3}.
\end{eqnarray}
We introduced the relative entropy
\begin{eqnarray}
\mathbb{D}_{kk'}^{(\nu)} \equiv k_b \sum_{i_k,j_{k'}} f_{i_k,j_{k'}}^{(\nu)} \ln \frac{f_{i_k,j_{k'}}^{(\nu)}}{f_{j_{k'},i_k}^{(\nu)}} \label{Dterm},
\end{eqnarray}
expressed in terms of conditional probabilities that, if a jump due to reservoir $\nu$ occurs from $k'$ to $k$, it is a jump from $j_{k'}$ to $i_k$:
\begin{eqnarray}
f_{i_k,j_{k'}}^{(\nu)} = \frac{W_{i_k,j_{k'}}^{(\nu)} \mathbb{P}_{j_{k'}}}{V_{kk'}^{(\nu)}} \label{DefF}
\end{eqnarray}
Normalization reads $\sum_{i_k,j_{k'}} f_{i_k,j_{k'}}^{(\nu)}=1$. This means that $\dot{S}^{(3)}_{\rm \bold i}$ can be interpreted as the entropy production contribution arising from the randomness associated to the different ways in which a given transition between macrostates can occur. Using the inequality $\ln x \leq x-1$, we prove that $\mathbb{D}_{kk'}^{(\nu)} \geq 0$. Therefore, the three contributions to the entropy production (\ref{EP}) are separately non-negative
\begin{eqnarray}
\dot{S}^{(1)}_{\rm \bold i} \geq 0 \  \ , \  \ \dot{S}^{(2)}_{\rm \bold i} \geq 0 \  \ , \  \ \dot{S}^{(3)}_{\rm \bold i} \geq 0 .\label{PositEPpieces}
\end{eqnarray}
This leads to the important result that neglecting any of these contributions will always imply an underestimation of entropy production. We note that if only single transitions connect pairs of macrostates via a given reservoir, $f_{i_k,j_{k'}}^{(\nu)}=1$, and therefore $\mathbb{D}_{kk'}^{(\nu)}=0$ and $\dot{S}^{(3)}_{\rm \bold i}=0$.

We finally separate the entropy flow (\ref{EntropyFlow}) as
\begin{eqnarray}
\dot{S}_{\rm \bold e} = \dot{S}^{(1)}_{\rm \bold e} + \dot{S}^{(2)}_{\rm \bold e} + \dot{S}^{(3)}_{\rm \bold e} .\label{EF}
\end{eqnarray}
Proceeding as before, we introduced the entropy flow arising from the dynamics between structure-less macrostates 
\begin{eqnarray}
\dot{S}^{(1)}_{\rm \bold e} = - k_b \sum_{\nu,k,k'} V_{kk'}^{(\nu)} P_{k'} \ln \frac{V_{kk'}^{(\nu)}}{V_{k'k}^{(\nu)}} .\label{EF1}
\end{eqnarray}
The entropy flow arising from the dynamics within macrostates is given by
\begin{eqnarray}
\dot{S}^{(2)}_{\rm \bold e} 
&=&  \sum_{k} P_{k} \bigg( \sum_{i_k} \dot{s}_{i_k} \mathbb{P}_{i_k} \label{EF2}\\
&& + \sum_{\nu,i_k,j_k} W_{i_k,j_k}^{(\nu)} \mathbb{P}_{j_k}  \big(s_{i_k}-s_{j_k} - k_b \ln \frac{W_{i_k,j_k}^{(\nu)}}{W_{j_k,i_k}^{(\nu)}} \big) \bigg) \nonumber \\ 
&=& \sum_k P_k \bigg( \sum_{i_k} \dot{s}_{i_k} \mathbb{P}_{i_k} + \sum_{\nu} \frac{\dot{\mathbb{Q}}^{(\nu)}(k)}{T^{(\nu)}} \bigg) \nonumber .
\end{eqnarray}
We introduced the heat conditional on being on the macrostate $k$ 
\begin{eqnarray}
\dot{\mathbb{Q}}^{(\nu)}(k)= \mathbb{I}_E^{(\nu)}(k) -\mu^{(\nu)} \mathbb{I}_M^{(\nu)}(k)
\end{eqnarray}
and the associated conditional energy and matter currents
\begin{eqnarray}
&&\mathbb{I}_E^{(\nu)}(k) = \sum_{i_k,j_k} W_{i_k,j_k}^{(\nu)} \mathbb{P}_{j_k} \big( \epsilon_{i_k} - \epsilon_{j_k} \big) \\ 
&&\mathbb{I}_M^{(\nu)}(k) = \sum_{i_k,j_k} W_{i_k,j_k}^{(\nu)} \mathbb{P}_{j_k} \big( n_{i_k} - n_{j_k} \big) \nonumber.
\end{eqnarray}
The last contribution to the entropy flow arising from the dynamics between microstates belonging to the different macrostates is given by
\begin{eqnarray}
\dot{S}^{(3)}_{\rm \bold e} &=&\sum_{\nu,k,k'(\neq k)} V_{kk'}^{(\nu)} p_{k'} \label{EF3}\\
&&\hspace{0cm} \bigg( \sum_{\nu,i_k,j_{k'}} f_{i_k,j_{k'}}^{(\nu)} \big(s_{i_k}-s_{j_{k'}} 
- k_b \ln \frac{f_{j_{k'},i_k}^{(\nu)} \mathbb{P}_{j_{k'}}}{f_{i_k,j_{k'}}^{(\nu)} \mathbb{P}_{i_{k}}} \big) \bigg) \nonumber .
\end{eqnarray}

It is essential to realize that the separation of the entropy balance into three contributions (the first associated to the dynamics of structure-less macrostates, the second to the internal dynamics of the macrostates and the third to the various ways in which transitions between macrostates can occur) is consistent in the sense that
\begin{eqnarray}
\dot{S}^{(1)} &=& \dot{S}^{(1)}_{\rm \bold i} + \dot{S}^{(1)}_{\rm \bold e} \label{Ebal1}\\
\dot{S}^{(2)} &=& \dot{S}^{(2)}_{\rm \bold i} + \dot{S}^{(2)}_{\rm \bold e} \label{Ebal2}\\
\dot{S}^{(3)} &=& \dot{S}^{(3)}_{\rm \bold i} + \dot{S}^{(3)}_{\rm \bold e} \label{Ebal3} .
\end{eqnarray}

We are now going to consider different limiting cases. We start by considering a non-driven system which has reached its steady state (i.e. all $\dot{p}_i=0$ and therefore $\dot{P}_k=0$ as well as $\dot{\mathbb{P}}_{i_k}=0$). In this case, we verify that 
\begin{eqnarray} 
&&\dot{S}=\dot{S}^{(1)}=\dot{S}^{(2)}+\dot{S}^{(3)}=0 \\
&&\dot{S}_{\rm \bold i}=-\dot{S}_{\rm \bold e} \ \ , \ \ \dot{S}^{(1)}_{\rm \bold i}=-\dot{S}^{(1)}_{\rm \bold e} \\
&&\dot{S}^{(2)}_{\rm \bold i}+\dot{S}^{(3)}_{\rm \bold i}=-\dot{S}^{(2)}_{\rm \bold e}-\dot{S}^{(3)}_{\rm \bold e}
\end{eqnarray}
If the time-scale separation described in section \ref{AdiabElim} is justified, the steady state macroscopic probabilities $P_k$ can be approximated by $P_k^{\rm st}$ and the conditional probabilities $\mathbb{P}_{i_k}$ by $\mathbb{P}_{i_k}^{\rm st}$, in all the expressions obtained above. In the next section, we will test the validity of this result on a specific model system.

We now turn to a different situation and consider a driven system. We assume that all transitions within a given macrostate $k$ are due to a single reservoir $\nu$. We also assume that the driving is slower than $\tau_{\rm mic}$, the time needed for $\mathbb{P}_{i_{k}}$ to equilibrate to $\mathbb{P}_{i_{k}}^{\rm eq}$. This means that for times longer than $\tau_{\rm mic}$, detailed balance will be satisfied within macrostates (\ref{DBink}). We find that
\begin{eqnarray} 
&&\dot{S}^{(2)}=\dot{S}^{(2)}_{\rm \bold e}=\dot{\lambda} \sum_k \partial_{\lambda} \mathbb{S}^{\rm eq}(k) P_{k} \\ 
&&\dot{S}^{(2)}_{\rm \bold i}=\dot{\mathbb{Q}}^{(\nu)}(k)=\mathbb{I}_E^{(\nu)}(k)=\mathbb{I}_M^{(\nu)}(k)=0 
\end{eqnarray}
Until now, we did not assume that the macrostates were thermalizing due to the same reservoirs, i.e. the $\mathbb{P}_{i_{k}}^{\rm eq}$ could correspond to different temperatures or chemical potentials for different $k$'s. To proceed, we now do so. By replacing $\mathbb{P}_{i_{k}}$ by $\mathbb{P}_{i_{k}}^{\rm eq}$ and $V_{kk'}^{(\nu)}$ by $V_{kk'}^{(\nu)\rm eq}$ in (\ref{DefF}) and (\ref{EP1}), we verify that 
\begin{eqnarray}
f_{i_k,j_{k'}}^{\rm eq} = f_{j_{k'},i_k}^{\rm eq}. \label{samef}
\end{eqnarray}
This also leads to 
\begin{eqnarray}
\dot{S}^{(3)}_{\rm \bold i}=\mathbb{D}_{kk'}^{(\nu)\rm eq}=0,
\end{eqnarray}
and as a result to
\begin{eqnarray}
\dot{S}^{(3)}=\dot{S}^{(3)}_{\rm \bold e}=\sum_{k} \mathbb{S}^{\rm eq}(k) \dot{P}_k.
\end{eqnarray}
In this case, we verify that summing up the three entropy contributions, we recover (\ref{EntMeso1EvolEq}):
\begin{eqnarray}
\hspace{-0.2cm} \dot{S} = \sum_k [ \mathbb{S}^{\rm eq}(k) - k_b \ln P_k] \dot{P_k} + \dot{\lambda} \sum_k \partial_{\lambda} \mathbb{S}^{\rm eq}(k) P_k . \nonumber
\end{eqnarray}
However, for entropy production and entropy flow we find
\begin{eqnarray}
\dot{S}_{\rm \bold i} = k_b \sum_{\nu,k,k'} V_{kk'}^{(\nu){\rm eq}} P_{k'} \ln \frac{V_{kk'}^{(\nu){\rm eq}} P_{k'}}{V_{k'k}^{(\nu){\rm eq}} P_{k}} \geq 0 ,\label{EntprodMeso1EvolEqGen}
\end{eqnarray}
and 
\begin{eqnarray}
\dot{S}_{\rm \bold e} &=& \sum_{\nu,k,k'} V_{kk'}^{(\nu){\rm eq}} P_{k'} \big(\mathbb{S}^{\rm eq}(k) - k_b \ln \frac{V_{kk'}^{(\nu){\rm eq}}}{V_{k'k}^{(\nu){\rm eq}}} \big) \label{EntflowMeso1EvolEqGen} \\
&&+ \dot{\lambda} \sum_k \partial_{\lambda} \mathbb{S}^{\rm eq}(k) P_k \nonumber,
\end{eqnarray}
which generalize (\ref{EntflowMeso1EvolEq}) and (\ref{EntprodMeso1EvolEq}) to cases where transitions between different macrostates can be due to different reservoirs. By assuming an equilibration within macrostates corresponding to the same reservoir (same temperature and chemical potential), we recovered at the macrostate level the most general formulation of stochastic thermodynamics presented at the microstate level in section \ref{Basics}.  

\section{Application to double coupled dots}\label{Applic}

\begin{figure}[h]
\rotatebox{0}{\scalebox{0.8}{\includegraphics{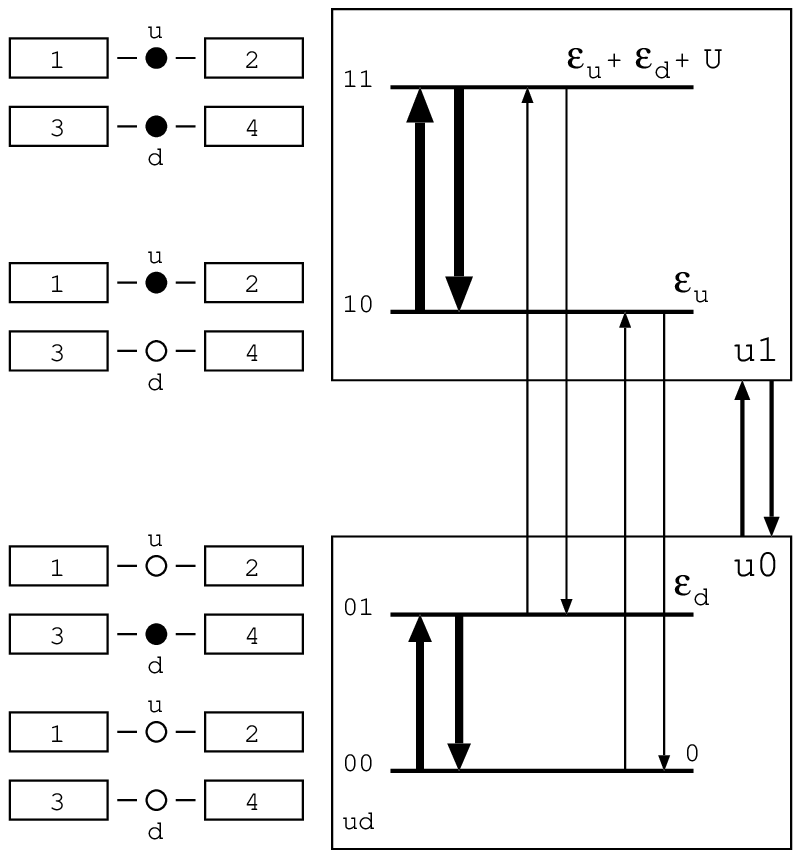}}}
\caption{Schematic illustration the coupled double dot model. Transitions in dot $d$ are faster than those in dot $u$. The coarse-graining procedure consists in grouping the four ``microscopic" states $ud=00,10,01,11$ into two ``macroscopic states" $u0$ and $u1$.}
\label{plot1}
\end{figure}

\begin{figure}[h]
\rotatebox{0}{\scalebox{0.6}{\includegraphics{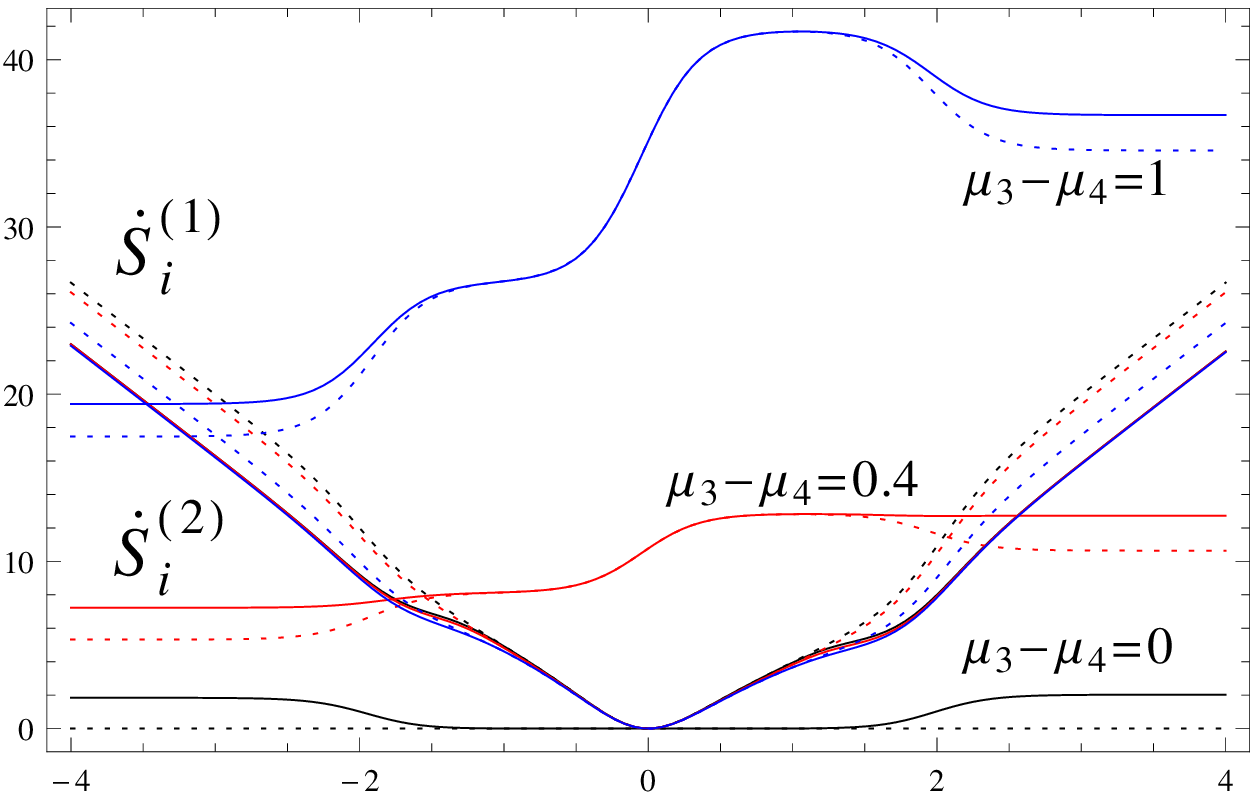}}}
\rotatebox{0}{\scalebox{0.6}{\includegraphics{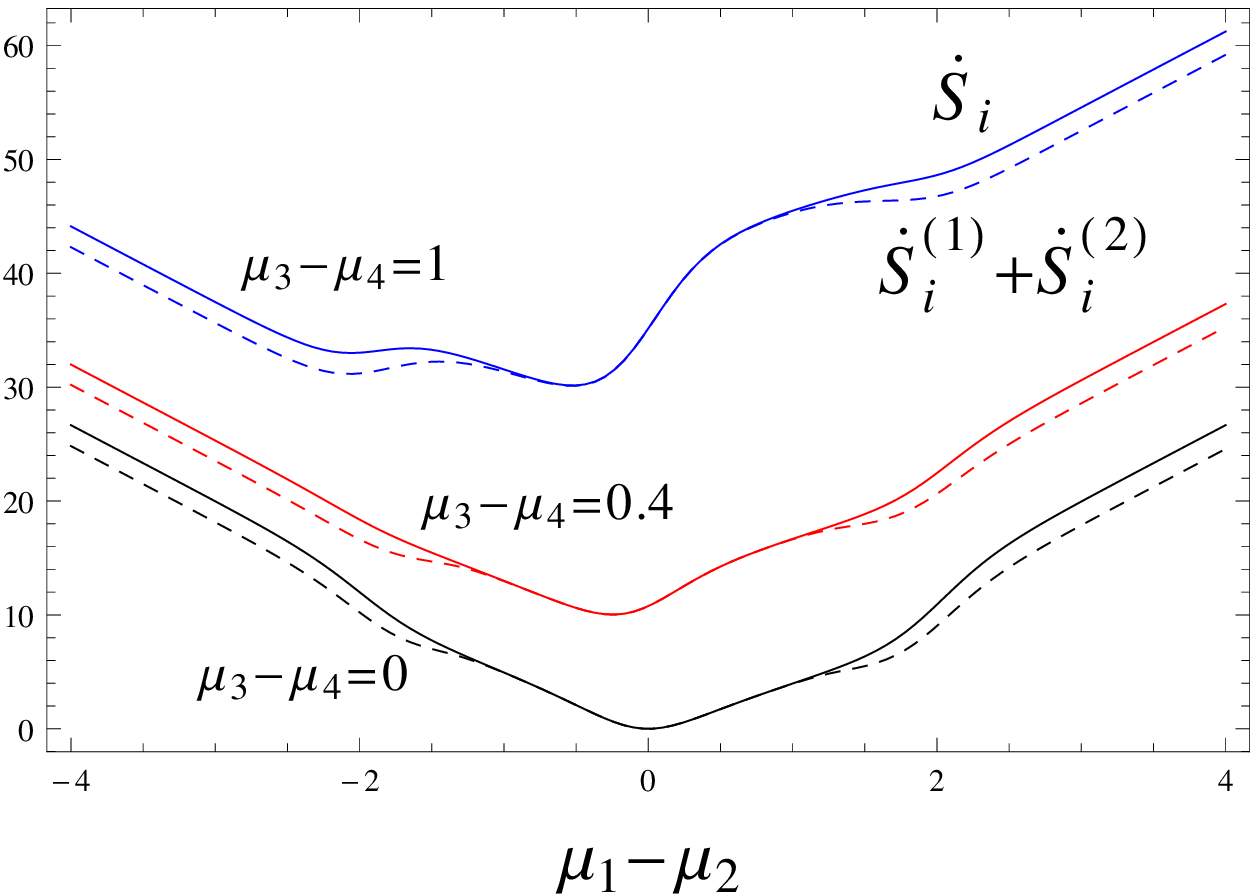}}}
\caption{(Color Online) Upper plot: $\dot{S}^{(1)}_{\rm \bold i}$ and $\dot{S}^{(2)}_{\rm \bold i}$ as a function of the bias $\mu_1-\mu_2$ in the slow dot $u$, for three different values of the bias $\mu_3-\mu_4$ in the fast dot $d$. The full line is the exact result and the dotted line is the approximate solution based on timescale separation. Lower plot: $\dot{S}_{\rm \bold i}$ (full line) and $\dot{S}^{(1)}_{\rm \bold i}+\dot{S}^{(2)}_{\rm \bold i}$ (dashed line). The difference between the full and dotted curve measures $\dot{S}^{(3)}_{\rm \bold i}$.
$\Gamma_1=1$, $\Gamma_2=2$, $\Gamma_3=11$, $\Gamma_4=10$.}
\label{plotA&B}
\end{figure}

\begin{figure}[h]
\rotatebox{0}{\scalebox{0.55}{\includegraphics{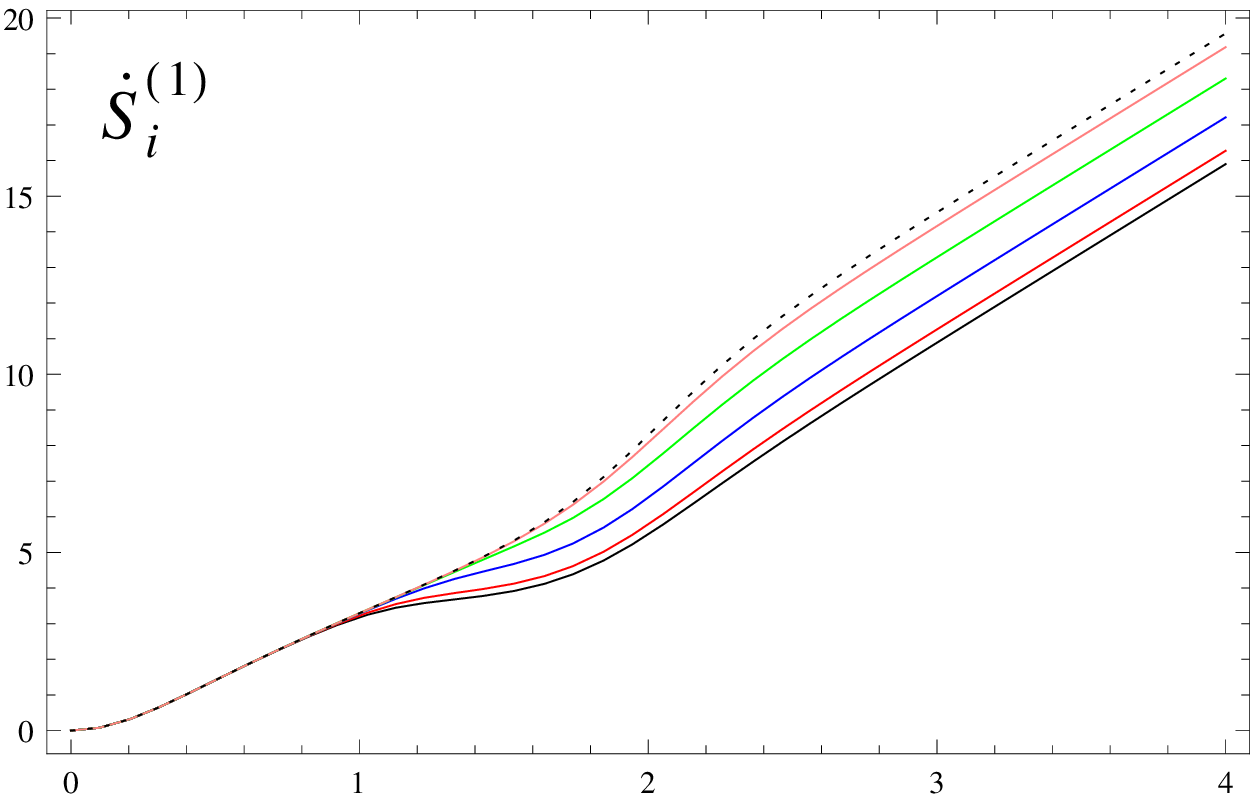}}}
\rotatebox{0}{\scalebox{0.555}{\includegraphics{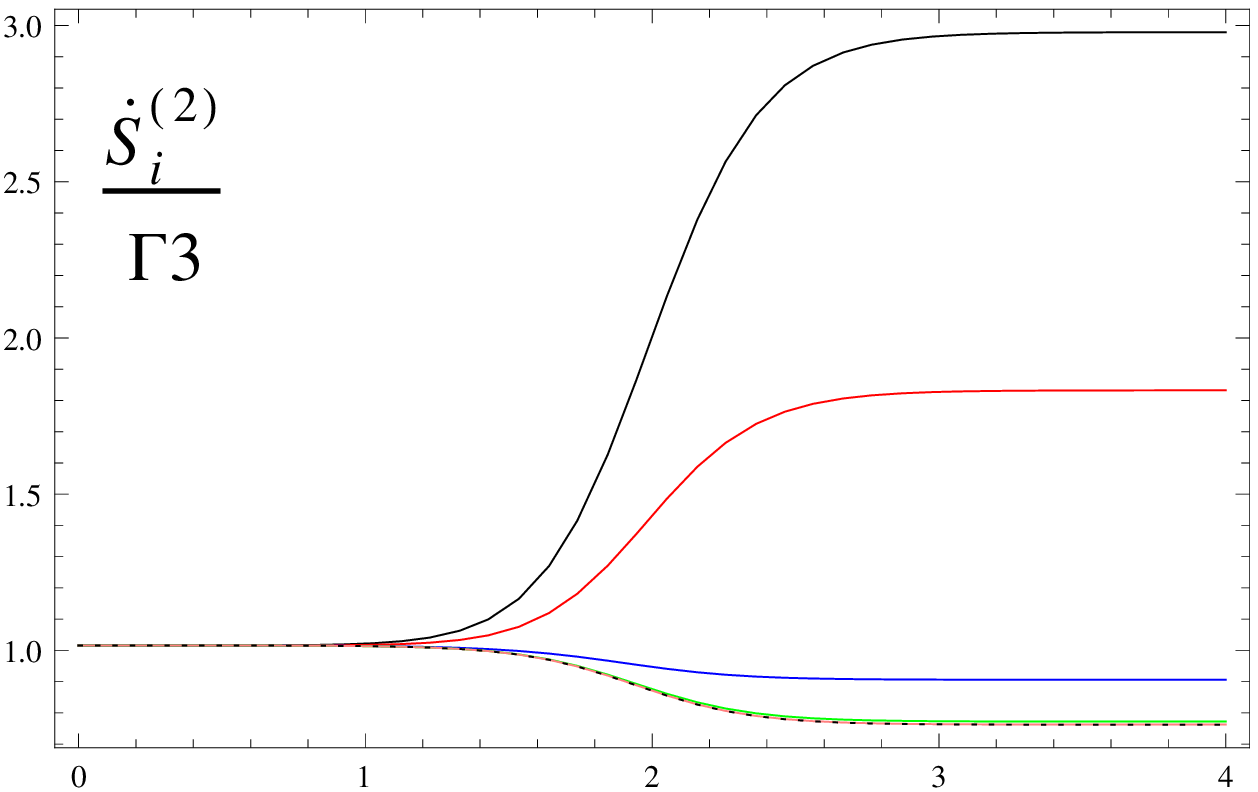}}}
\rotatebox{0}{\scalebox{0.55}{\includegraphics{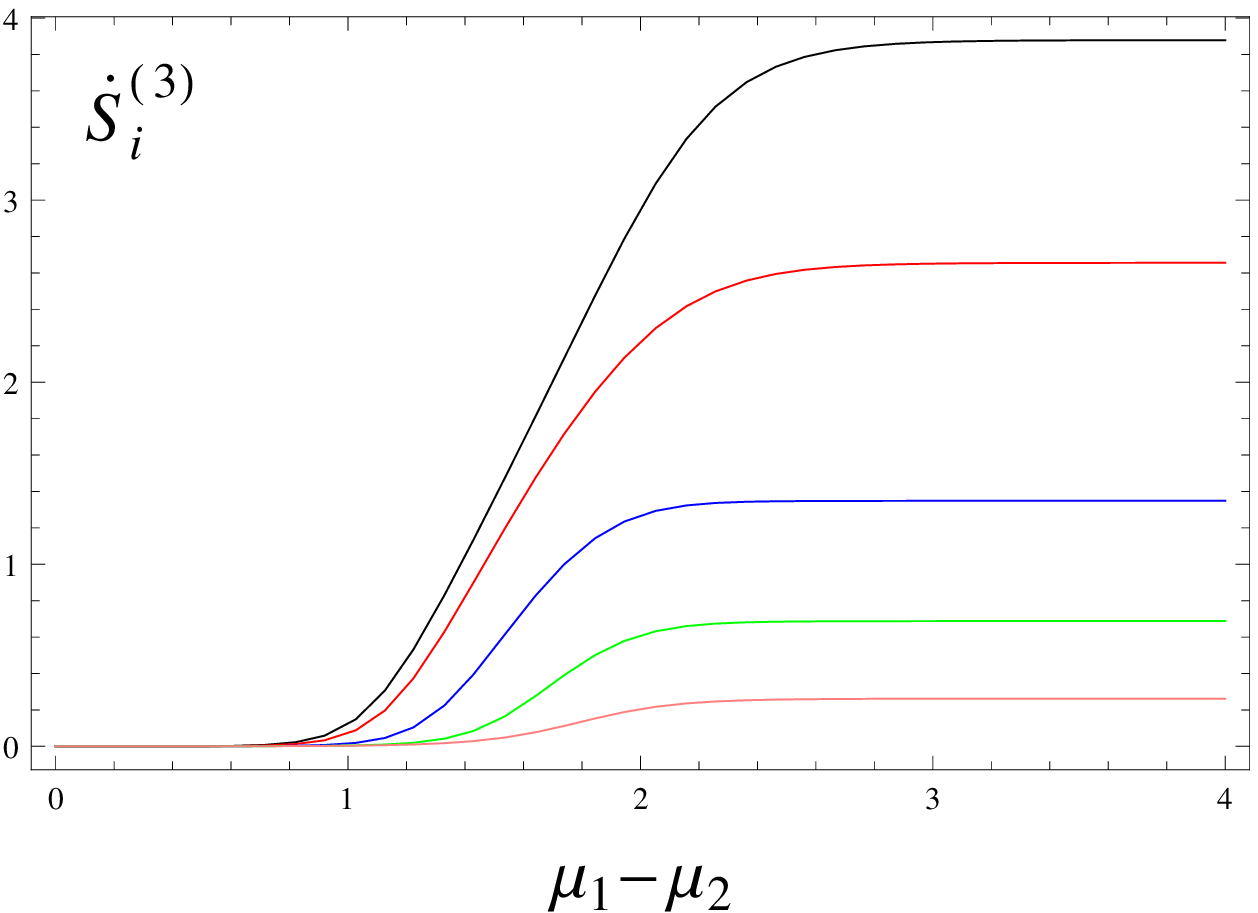}}}
\caption{(Color Online) On the first (resp. second and third) plot, the full lines from below to above (resp. from above to below) correspond to $\Gamma_3=\Gamma_4=0.1 ({\rm black}),$ $1 ({\rm red}),$ $10 ({\rm blue}),$ $100 ({\rm green}),$ $1000 ({\rm pink})$. The dotted line is the approximate solution based on timescale separation. $\mu_3-\mu_4=0.4$ and $\Gamma_1=\Gamma_2=1$.}
\label{plotF&G&H}
\end{figure}

To illustrate the different contribution to entropy production, we consider a model of two coupled single level quantum dots which was previously studied in Refs. \cite{ButtikerSanchez10PRL, BrandesSchallerPRB10, EspositoCuetaraGaspard11} and which is illustrated on Fig. \ref{plot1}. An electron in the lower dot $d$ (resp. upper dot $u$) has an energy $\epsilon_d$ (resp. $\epsilon_u$). An electron-electron interaction denoted by $U$ takes place when both dots are occupied. The four microscopic states have no internal structure and thus no internal entropy. They are labeled by $ud=00,10,01,11$ and have corresponding energies $\epsilon_{00}=0$, $\epsilon_{01}=\epsilon_d$, $\epsilon_{10}=\epsilon_u$ and $\epsilon_{10}=\epsilon_u+\epsilon_s+U$ and number of particles $n_{00}=0$, $n_{01}=1$, $n_{10}=1$ and $n_{11}=2$. The four reservoirs causing transitions between these states are labeled by $\nu=1,2,3,4$. Transfers of electrons in and out of dot $u$ (resp. $d$) are caused by reservoir $1$ or $2$ (resp. $3$ or $4$). No electron transfer can occur between the two dots. The rate equation describing the full dynamics is given by
\begin{widetext}
\begin{equation}{\label{e5}}
\begin{bmatrix}
\dot{p}_{11}\\
\dot{p}_{10}\\
\dot{p}_{01}\\
\dot{p}_{00}
\end{bmatrix}
=
\begin{bmatrix}
-W_{01,11}-W_{10,11} & W_{11,10}             & W_{11,01}            & 0 \\
W_{10,11}            & -W_{00,10}-W_{11,10}  & 0                    & W_{10,00} \\
W_{01,11}            & 0                     & -W_{00,01}-W_{11,01} & W_{01,00} \\
0                    & W_{00,10}             & W_{00,01}            & -W_{01,00}-W_{10,00} \\
\end{bmatrix}
\begin{bmatrix}
p_{11}\\
p_{10}\\
p_{01}\\
p_{00}
\end{bmatrix},
\end{equation}
where the detailed form of each of the rates is given  
\begin{equation}
\begin{array}{lll}
W_{11,01}=W_{11,01}^{(1)} + W_{11,01}^{(2)} \ \ {\rm where} 
\ \ &W_{11,01}^{(1)}= \Gamma^{(1)} f^{(1)}(\epsilon_u+U)              &\ \ , \ \  W_{11,01}^{(2)}= \Gamma^{(2)} f^{(2)}(\epsilon_u+U) \\
W_{01,11}=W_{01,11}^{(1)} + W_{01,11}^{(2)} \ \ {\rm where} 
\ \ &W_{01,11}^{(1)}= \Gamma^{(1)} \big(1-f^{(1)}(\epsilon_u+U) \big) &\ \ , \ \  W_{01,11}^{(2)}= \Gamma^{(2)} \big( 1- f^{(2)}(\epsilon_u+U) \big) \\
W_{10,00}=W_{10,00}^{(1)} + W_{10,00}^{(2)} \ \ {\rm where} 
\ \ &W_{10,00}^{(1)}= \Gamma^{(1)} f^{(1)}(\epsilon_u)                &\ \ , \ \  W_{10,00}^{(2)}= \Gamma^{(2)} f^{(2)}(\epsilon_u) \\
W_{00,10}=W_{00,10}^{(1)} + W_{00,10}^{(2)} \ \ {\rm where} 
\ \ &W_{00,10}^{(1)}= \Gamma^{(1)} \big(1-f^{(1)}(\epsilon_u) \big)   &\ \ , \ \  W_{00,10}^{(2)}= \Gamma^{(2)} \big( 1- f^{(2)}(\epsilon_u) \big) \\
W_{11,10}=W_{11,10}^{(3)} + W_{11,10}^{(4)} \ \ {\rm where} 
\ \ &W_{11,10}^{(3)}= \Gamma^{(3)} f^{(3)}(\epsilon_d+U)              &\ \ , \ \  W_{11,01}^{(4)}= \Gamma^{(4)} f^{(4)}(\epsilon_d+U) \\
W_{10,11}=W_{10,11}^{(3)} + W_{10,11}^{(4)} \ \ {\rm where} 
\ \ &W_{10,11}^{(3)}= \Gamma^{(3)} \big(1-f^{(3)}(\epsilon_d+U) \big) &\ \ , \ \  W_{10,11}^{(4)}= \Gamma^{(4)} \big( 1- f^{(4)}(\epsilon_d+U) \big) \\
W_{01,00}=W_{01,00}^{(3)} + W_{01,00}^{(4)} \ \ {\rm where} 
\ \ &W_{01,00}^{(3)}= \Gamma^{(3)} f^{(3)}(\epsilon_d)                &\ \ , \ \  W_{01,00}^{(4)}= \Gamma^{(4)} f^{(4)}(\epsilon_d)\\
W_{00,01}=W_{00,01}^{(3)} + W_{00,01}^{(4)} \ \ {\rm where} 
\ \ &W_{00,01}^{(3)}= \Gamma^{(3)} \big(1-f^{(3)}(\epsilon_d) \big)   &\ \ , \ \  W_{00,01}^{(4)}= \Gamma^{(4)} \big( 1- f^{(4)}(\epsilon_d) \big) 
\end{array}
\end{equation}
\end{widetext}
The Fermi distributions of the reservoirs are given by 
\begin{eqnarray}
f^{(\nu)}(\epsilon)= (1+\exp{\bigg(\frac{\epsilon-\mu^{\nu}}{k_b T}\bigg)})^{-1}
\end{eqnarray}
We consider reservoirs with same temperatures but different chemical potentials. We verify that the rates satisfy local detailed balance (\ref{LocalDB}). We propose a coarse-graining procedure motivated by the fact that when dot $d$ is faster than dot $u$, it can be used as a probe to detect the changes in the occupation of dot $u$ \cite{BrandesSchallerPRB10, EspositoCuetaraGaspard11}. In such situations only the state of dot $u$ is monitored. Therefore, the first macrostate, denoted $u0$, corresponds to dot $u$ empty and contains the two microstates $00$ and $01$. The second macrostate, denoted $u1$, corresponds to dot $u$ filled and contains the two microstates $10$ and $11$. These two macrostates are represented by a square on fig. \ref{plot1} which encapsulated the corresponding microscopic levels. We will only consider the steady state regime. We easily calculate the approximate steady-state conditional probabilities to be in a microstate corresponding to a given macrostate by neglecting transitions between macrostates   
\begin{eqnarray}
&&\mathbb{P}_{11|u1}^{\rm st}=W_{11,10}/(W_{11,10}+W_{10,11}) \\
&&\mathbb{P}_{10|u1}^{\rm st}=W_{10,11}/(W_{11,10}+W_{10,11}) \nonumber \\
&&\mathbb{P}_{01|u0}^{\rm st}=W_{01,00}/(W_{01,00}+W_{00,01}) \nonumber \\
&&\mathbb{P}_{00|u0}^{\rm st}=W_{00,01}/(W_{01,00}+W_{00,01}) \nonumber.
\end{eqnarray}
Using these, we can construct the approximate rate matrix at the macrostate level 
\begin{eqnarray}
&&V_{u0,u1}^{(1,2)\rm st} = W_{00,10}^{(1,2)} \mathbb{P}_{10|u1}^{\rm st} + W_{01,11}^{(1,2)} \mathbb{P}_{11|u1}^{\rm st} \\
&&V_{u1,u0}^{(1,2)\rm st} = W_{10,00}^{(1,2)} \mathbb{P}_{00|u0}^{\rm st} + W_{11,01}^{(1,2)} \mathbb{P}_{01|u0}^{\rm st} \nonumber.
\end{eqnarray}
From these we can calculate the corresponding approximate macroscopic probabilities 
\begin{eqnarray}
P_{u1}^{\rm st}=V_{u1,u0}^{\rm st}/(V_{u1,u0}^{\rm st}+V_{u0,u1}^{\rm st}) \\
P_{u0}^{\rm st}=V_{u0,u1}^{\rm st}/(V_{u1,u0}^{\rm st}+V_{u0,u1}^{\rm st}) \nonumber.
\end{eqnarray}
Using these result, we can easily calculate the approximate form of the different contributions to entropy production, $\dot{S}^{(1)}_{\rm \bold i}$, $\dot{S}^{(2)}_{\rm \bold i}$, $\dot{S}^{(3)}_{\rm \bold i}$ presented in section \ref{MesoMutiResSec}. To asses the accuracy of the approximate solutions based on time-scale separation, we compare them with the exact resuts obtained from numerical simulation in Fig. \ref{plotA&B} and Fig. \ref{plotF&G&H}. We always choose $k_b T=0.1$, $\epsilon_d=\epsilon_u=U=1$ and the gate voltages as $\mu_1+\mu_2=2$ and $\mu_3+\mu_4=2$.
In Fig. \ref{plotA&B}, we see that an increase in the bias inside the macro-levels, $\mu_3-\mu_4$, obviously increases $\dot{S}^{(2)}_{\rm \bold i}$ but leaves the macro-level entropy production, $\dot{S}^{(1)}_{\rm \bold i}$, almost unaffected. The converse is not true. Indeed, $\dot{S}^{(1)}_{\rm \bold i}$ is obviously affected by an increase of the bias $\mu_1-\mu_2$, but $\dot{S}^{(2)}_{\rm \bold i}$ is also non-trivially affected. This shows that dissipation in the detector (quantum dot $d$) is sensible to the state of the detected quantum dot $u$, while dissipation in the detected dot is relatively unaffected by the presence of the detector. We also note that $\dot{S}^{(3)}_{\rm \bold i}$ always tends to remain relatively small. This is because transitions between macrostates can only occur in two different ways in this model. We also note that the approximate solutions are quite accurate despite a very moderate time scale separation. For higher separation (e.g. $\Gamma_3=110$, $\Gamma_4=100$), the discrepancy is so small that it is not visible anymore (not shown).
In Fig. \ref{plotF&G&H}, we see that as long as the bias in dot $u$ is lower than in dot $d$ (i.e. microlevels are further away from equilibrium than macrolevels), the approximate contributions to entropy production fit very well with the exact ones. However, as the bias in dot $u$ becomes larger than in dot $d$, keeping a low discrepancy between the exact and approximate contributions to entropy production requires an increasing time scale separation between $u$ and $d$. 

\section{Conclusions}\label{Conc}

To the best of our knowledge, this paper presents the most complete formulation of stochastic thermodynamics. The theory is developed to describes open systems, in contact to multiple reservoirs, and externally driven by time-dependent forces. It also accounts for situations where the system states have an internal structure which may be affected by the external driving. By performing a coarse-graining of the system ``microstates" in terms of ``macrostates", we analyze how the thermodynamic description of the system, and in particular its entropy balance, gets affected. We find for example that entropy production is made of three separately positive contributions. The first arises from the dynamics between macrolevels. The second is a macrostate ensemble average of the entropy production arising from within each macrostate. The last one results from the multiple ways in which transitions between macrostates can occur. We also identify the precise conditions under which the mathematical structure of the theory at the microlevel is recovered after the coarse-graining procedure at the macrolevel. This a posteriori clarifies the implicit assumptions made when describing a system using stochastic thermodynamics. We now repeat these central assumptions. Stochastic thermodynamics describes an open system in term of probabilities to find the system in certain states. These probabilities can be arbitrary far from equilibrium and evolve according to a Markovian stochastic dynamics. The system states may have an internal structure (which might even change under the effect of an external force) but which always remains at equilibrium so that a well defined energy, particle number, and entropy, can be associated to them. Transitions between these states are induced by reservoirs which each adds an independent contribution to the rate matrix. These separate contributions satisfy local detailed balance because reservoirs are sets of equilibrated degrees of freedom fully characterized by a temperature and a chemical potential. This paper shows that the internal consistency of stochastic thermodynamics is remarkable. In view of its already many successful applications, we are convinced that this theory will become an essential tool for the study of small devices operating far-from-equilibrium. 

\section*{Acknowledgments}

M. E. is supported by the Belgian Federal Government (IAP project ``NOSY") and by the European Union Seventh Framework Programme (FP7/2007-2013) under grant agreement 256251.

\vskip 10pt
\appendix
\section{Justifying adiabatic elimination} \label{AppAdiabElimRigor}

We consider a rate matrix which has a structure such that transitions between the microstates involving different macrostates are a factor $\gamma^{-1}$ slower than those involving the same macrostates: 
\begin{eqnarray}
W_{i_k j_{k'}}^{(\nu)} = W_{i_k j_k}^{(\nu)} \delta_{k,k'} + \gamma^{-1} W_{i_k j_{k'}}^{(\nu)} (1-\delta_{k,k'}) \label{SplitOfRate}.
\end{eqnarray}
We therefore search for solutions of the form 
\begin{eqnarray}
&& P_k = P_k^{(0)} + \gamma^{-1} P_k^{(1)} + \gamma^{-2} P_k^{(2)} + \hdots \label{ExpMacProb} \\
&& \mathbb{P}_{i_k} = \mathbb{P}_{i_k}^{(0)} + \gamma^{-1} \mathbb{P}_{i_k}^{(1)} + \gamma^{-2} \mathbb{P}_{i_k}^{(2)} + \hdots \label{ExpMicProb}.
\end{eqnarray}
Normalization implies that $\sum_{k} P_k^{(0)}=1$, $\sum_{k} P_k^{(n)}=0$ for $n \geq 1$, $\sum_{i_k} \mathbb{P}_{i_k}^{(0)}=1$ and $\sum_{i_k} \mathbb{P}_{i_k}^{(n)}=0$ for $n \geq 1$. We start by using $\sum_k V_{kk'} = 0$, which implies that $V_{k'k'} = -\sum_{k (\neq k')} V_{kk'}$. At order $\gamma^{0}$ and $\gamma^{-1}$, we find 
\begin{eqnarray}
&&\sum_{i_{k'},j_{k'},\nu} W_{i_{k'} j_{k'}}^{(\nu)} \mathbb{P}_{j_{k'}}^{(0)} = 0 \label{EffRateOrder0} \\
&&\sum_{i_{k'},j_{k'},\nu} W_{i_{k'} j_{k'}}^{(\nu)} \mathbb{P}_{j_{k'}}^{(1)} = - \sum_{k (\neq k')} \sum_{i_k,j_{k'},\nu} W_{i_k j_{k'}}^{(\nu)} \mathbb{P}_{j_{k'}}^{(0)} \label{EffRateOrder1}.
\end{eqnarray}
Using (\ref{ExpMacProb}), (\ref{ExpMicProb}) and (\ref{SplitOfRate}) in (\ref{ComplDyn}), we find at order $\gamma^{0}$
\begin{eqnarray}
\dot{P}_k^{(0)} \mathbb{P}_{i_k}^{(0)} + P_{k}^{(0)} \dot{\mathbb{P}}_{i_k}^{(0)} = P_{k}^{(0)} \sum_{j_k,\nu} W_{i_k j_k}^{(\nu)} \mathbb{P}_{j_k}^{(0)} \label{ComplDynOrder0}.
\end{eqnarray}
Summing (\ref{ComplDynOrder0}) over $i_k$ and using (\ref{EffRateOrder0}), we get that
\begin{eqnarray}
\dot{P}_k^{(0)} = 0 \label{Pk0}.
\end{eqnarray}
Turning back to (\ref{ComplDynOrder0}), we thus get
\begin{eqnarray}
\dot{\mathbb{P}}_{i_k}^{(0)} = \sum_{j_k,\nu} W_{i_k j_k}^{(\nu)} \mathbb{P}_{j_k}^{(0)} \label{ComplDynOrder0bis}.
\end{eqnarray}
Using (\ref{Pk0}) and (\ref{ComplDynOrder0bis}), we find that at order $\gamma^{-1}$, (\ref{ComplDyn}) reads 
\begin{eqnarray}
\dot{P}_k^{(1)} \mathbb{P}_{i_k}^{(0)} + P_{k}^{(0)} \dot{\mathbb{P}}_{i_k}^{(1)} &=&  P_{k}^{(0)} \sum_{j_k,\nu} W_{i_k j_k}^{(\nu)} \mathbb{P}_{j_{k}}^{(1)} \label{ComplDynOrder1}\\
&&+ \sum_{k'(\neq k)} P_{k'}^{(0)} \sum_{j_{k'},\nu} W_{i_k j_{k'}}^{(\nu)} \mathbb{P}_{j_{k'}}^{(0)} \nonumber .
\end{eqnarray}
Summing (\ref{ComplDynOrder1}) over $i_k$ and using (\ref{EffRateOrder1}), we get
\begin{eqnarray}
\dot{P}_k^{(1)} = \sum_{k'(\neq k)} \big( V_{kk'}^{(\nu)(0)} P_{k'}^{(0)} - V_{k'k}^{(\nu)(0)} P_{k}^{(0)} \big) \label{MEmacrostatesOrder0} ,
\end{eqnarray}
where 
\begin{eqnarray}
V_{kk'}^{(\nu)(0)} = \sum_{i_k,j_{k'}} W_{i_k j_{k'}}^{(\nu)} \mathbb{P}_{j_{k'}}^{(0)} \label{EffMacRateOrder0} .
\end{eqnarray}
Since we limit ourself to order $\gamma^{-1}$ and neglect order $\gamma^{-2}$ and higher corrections, we can close (\ref{MEmacrostatesOrder0}) by writing it as 
\begin{eqnarray}
\hspace{-0.5cm} \dot{P}_k^{(0+1)} = \sum_{k'(\neq k)} \big( V_{kk'}^{(\nu)(0)} P_{k'}^{(0+1)} - V_{k'k}^{(\nu)(0)} P_{k}^{(0+1)} \big) \label{MEmacrostatesOrder0b} ,
\end{eqnarray}
where $P_{k}^{(0+1)}=P_{k}^{(0)} + \gamma^{-1} P_{k}^{(1)}$.


%

\end{document}